\documentclass{article}

\PassOptionsToPackage{square,numbers}{natbib}
\usepackage[final]{tackling_climate_workshop_style}

\usepackage[utf8]{inputenc} % allow utf-8 input
\usepackage[T1]{fontenc}    % use 8-bit T1 fonts
\usepackage{hyperref}       % hyperlinks
\usepackage{url}            % simple URL typesetting
\usepackage{booktabs}       % professional-quality tables
\usepackage{amsfonts}       % blackboard math symbols
\usepackage{nicefrac}       % compact symbols for 1/2, etc.
\usepackage{microtype}      % microtypography
\usepackage[pdftex]{graphicx}

\title{Deep Learning for Rapid Landslide Detection using Synthetic Aperture Radar (SAR) Datacubes}

\author{
  Vanessa Boehm \thanks{Authors contributed equally.} \\
  University of California Berkeley \\
  United States \\
  \And
  Wei Ji Leong $^\ast$\\
  The Ohio State University \\
  United States \\
  \And
      Ragini Bal Mahesh $^\ast$ \\
  German Aerospace Center DLR \\
  Germany \\
    \And
    Ioannis Prapas $^\ast$ \\
  University of Valencia, Spain \\
  National Observatory of Athens, Greece \\
    \And
  Edoardo Nemni \\
  United Nations Satellite Centre \\
  Switzerland \\
  \And
  Freddie Kalaitzis \\
  University of Oxford \\
  United Kingdom \\
  \texttt{} \\
  \And
  Siddha Ganju \\
  NVIDIA \\
  United States \\
  \And
  Raul Ramos-Pollan \\
  Universidad de Antioquia \\
  Colombia \\
}

\begin{document}

\maketitle

\begin{abstract}
With climate change predicted to increase the likelihood of landslide events, there is a growing need for rapid landslide detection technologies that help inform emergency responses.
Synthetic Aperture Radar (SAR) is a remote sensing technique that can provide measurements of affected areas independent of weather or lighting conditions.
Usage of SAR, however, is hindered by domain knowledge that is necessary for the pre-processing steps and its interpretation requires expert knowledge.
We provide simplified, pre-processed, machine-learning ready SAR datacubes for four globally located landslide events obtained from several Sentinel-1 satellite passes before and after a landslide triggering event together with segmentation maps of the landslides.
From this dataset, using the Hokkaido, Japan datacube, we study the feasibility of SAR-based landslide detection with supervised deep learning (DL).
Our results demonstrate that DL models can be used to detect landslides from SAR data, achieving an Area under the Precision-Recall curve exceeding 0.7.
We find that additional satellite visits enhance detection performance, but that early detection is possible when SAR data is combined with terrain information from a digital elevation model.
This can be especially useful for time-critical emergency interventions. Code is made publicly available at \url{https://github.com/iprapas/landslide-sar-unet}.
\end{abstract}

\section{Introduction}
According to the United Nations Office for Disaster Risk Reduction, landslides have affected 4.8 million people and caused 18,414 deaths between 1998-2017 \cite{wallemacq2018economic}. %\footnote{\url{https://www.preventionweb.net/files/61119_credeconomiclosses.pdf}}.
Rising temperatures and climate change are projected to worsen this situation~\cite{GarianoLandslidesChangingClimate2016,HuggelPhysicalImpactsClimate2012} with the increase of sustained droughts and intense rainfall events~\cite{AllenConstraintsFutureChanges2002,IngramIncreasesAll2016,BhatiaRecentIncreasesTropical2019}.
Given these predictions, there is a growing need for timely and accurate landslide assessment methods that can inform decision makers and emergency responders.
In the aftermath of a landslide, optical satellite imagery is commonly used to map the extent of the event. The fact that optical data is is often hindered by clouds and limited to day time observations, motivates the study of weather independent Synthetic Aperture Radar (SAR) as a monitoring technique.
%SAR is an active imagining sensor, which is independent of cloud coverage, time of the day, and weather conditions.

DL algorithms for computer vision can be used for SAR analysis if SAR data is represented as images (or georeferenced data arrays)~\cite{zhu_deep_2021}. This has been exploited for multiple use cases such as the detection of volcanic activity \cite{bountos2022learning1}, floods \cite{paul_flood_2021}, and landslides\cite{nava_rapid_2022}. The landslide change detection with DL can be performed in two ways: object-level \cite{nava_improving_2021} and pixel-level detection \cite{nava_rapid_2022}. As per the authors' best knowledge \cite{nava_rapid_2022,nava_improving_2021} are one of the first studies in landslide mapping using DL and SAR images where bi- and tri-temporal images have been used. Usage of multi-temporal SAR polarization modes has been demonstrated with traditional approaches, such as thresholding ~\cite{handwerger_generating_2022}, similar conclusion was reported in a DL study \cite{machielse2021landslide}. To achieve good accuracy, these methods require longer time series of SAR data (up to months after the event), which are unavailable in an emergency scenario.

%The feasibility of landslide detection with multi-temporal SAR polarization modes has been demonstrated with traditional approaches, such as thresholding~\cite{handwerger_generating_2022}. 
%To achieve good accuracy, these methods require longer time-series of SAR data (up to months after the event), which are unavailable in an emergency scenario. Deep learning algorithms for computer vision can be used for SAR analysis, if SAR data is represented as images (or georeferenced data arrays)~\cite{zhu_deep_2021}. This has been exploited for multiple use cases such as detection of volcanic activity \cite{bountos2022learning1}, floods \cite{paul_flood_2021} and landslides\cite{nava_rapid_2022}.

The development of DL algorithms for SAR analysis, however, is still hindered by its complex characteristics and pre-procesisng requirements.
To help overcome this barrier, we publicly release curated DL-ready datasets \cite{boehm_vanessa_2022_7248056} that includes SAR intensity and interferometry data before and after disaster events that triggered landslides.
Using part of this dataset, we further develop a supervised DL algorithm for landslide detection and demonstrate the feasibility of rapid detection using only one satellite pass, if SAR data is combined with terrain information.

\section{Data}
\begin{figure}[htbp]
  \centering
  \includegraphics[width=1.0\linewidth]{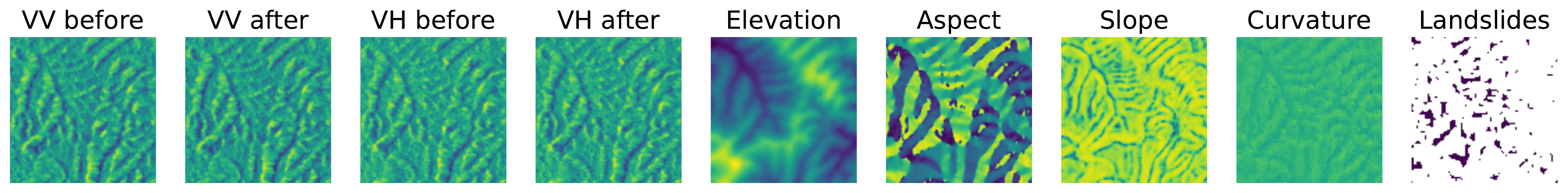}
  \caption{
    Sample data used to train the DL model.
    In each of panels 1-8, we show one channel of the input data, in the last panel we show the target landslide mask. 1-4: Polarized Synthetic Aperture Radar intensity images (VV and VH) from before and after the landslide trigger event; 4-8:
    Digital Elevation Model (DEM, elevation) and DEM-derived products (aspect, slope, curvature).
  }
  \label{fig:1}
\end{figure}

We produce analysis-ready datacubes \cite{boehm_vanessa_2022_7248056} with multiple layers of SAR time-series data and auxiliary geographic information for multiple landslide events.
To this end, we develop a data pipeline that processes SAR intensity and interferometric products from multiple sources and compile them together with a Digital Elevation Model (DEM) and DEM-derived products into a format that can be fed into DL models (see Fig.~\ref{fig:1}).

% A major obstacle to using SAR data is generally its complexity.
% The SAR backscatter signal is a backscattering signal, comprised of amplitude, phase and polarization information.
% The signal depends on many parameters including the orbital position of the satellite and its direction of motion, since SAR satellites are side-looking.
% As opposed to optical data, SAR data requires many expert-level preprocessing steps to be transformed into interpretable data products.

\begin{figure}[htbp]
  \centering
  \includegraphics[width=1.0\linewidth]{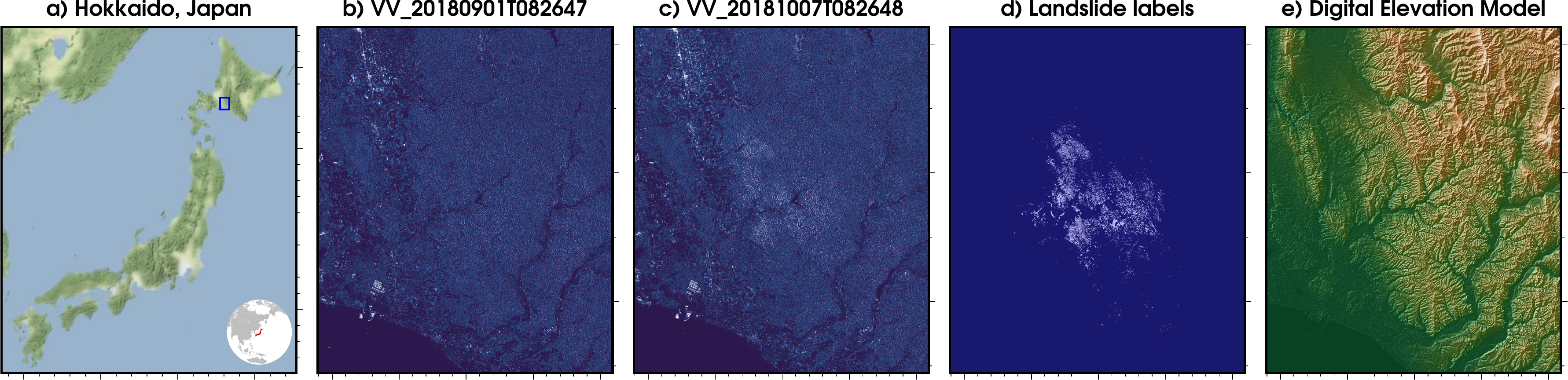}
  \caption{
    Location map of the Hokkaido, Japan study area (a, blue box) and data layers covering spatial area of about 236 km$^2$ (4398x5382 pixels).
    Two VV-channel SAR images before (b) and after (c) the earthquake are shown, together with the landslides labels (d) and SRTM DEM (e).
    The actual datacube covers a temporal period from 2018/07 to 2019/11, with SAR images in ascending mode only, consisting of 18 dual polarized SAR (VV and VH) timesteps.
    Although not used in this study, the datacube also contains InSAR channels (coherence and interferometry) between 16 pairs of consecutive acquisitions.
  }
  \label{fig:2_hokkaido_datacube}
\end{figure}

In this study, we focus our machine learning experiments on the datacube created for earthquake-triggered landslides over Hokkaidō, Japan in 2017 (Fig~\ref{fig:2_hokkaido_datacube}).
Additional datacubes are available for landslide events over Mt Talakmau, Indonesia (2022 earthquake), Kaikōura, New Zealand (2016 earthquake) and Puerto Rico (2017 Hurricane) (see Appendix~\ref{app:datacubes} for more details on datacube creation).

\section{Methodology}

\subsection{Experimental setup}
\begin{figure}[htbp]
  \centering
  \includegraphics[width=1.0\linewidth]{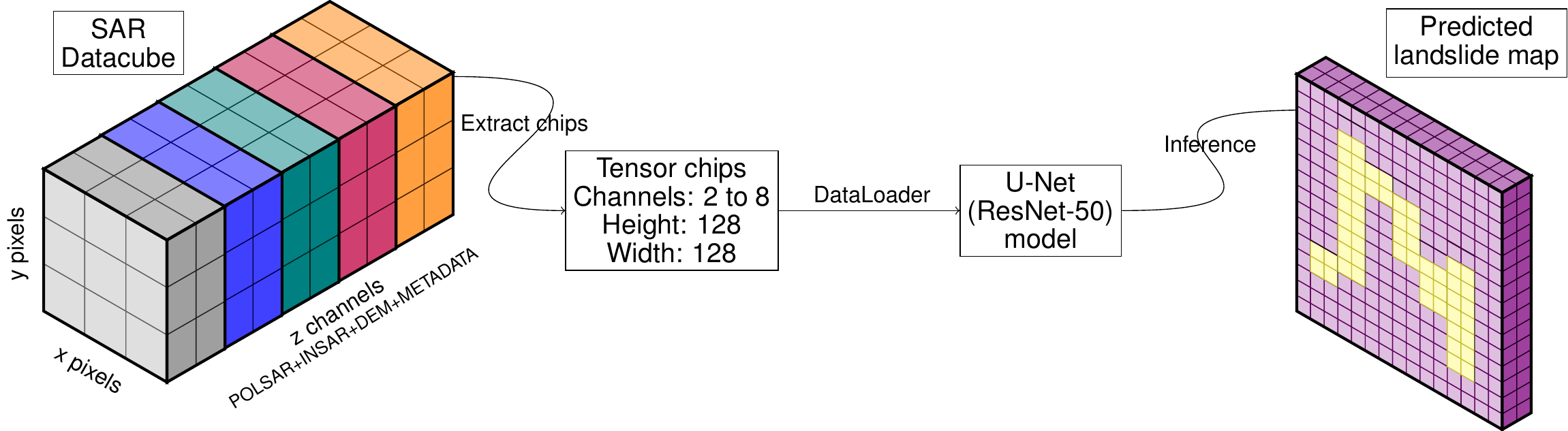}
  \caption{
    Machine Learning pipeline from the Synthetic Aperture Radar (SAR) datacube (left) to landslide map predictions (right).
    SAR time-series data and Digital Elevation Model (DEM) products are first extracted into 128x128 chips.
    Next, batches of chips are fed via a DataLoader into the U-Net model with a ResNet-50 encoder.
    The trained U-Net model is then used to do inference and produce landslide segmentation maps.
  }
  \label{fig:3}
\end{figure}

The machine learning pipeline is set up to solve a segmentation task (Fig.~\ref{fig:3}). We divide the Hokkaido datacube into data chips of size 128x128 pixels (each pixel is 10m$\times 10m$ in spatial resolution). 
Since the dataset is imbalanced, we reduce negative examples by keeping only data chips with one or more landslide pixels, similar to \citet{nava_improving_2021}.
These chips are split into training, and test sets with 216, and 61 chips respectively. Model performance is measured using the Area Under the Precision Recall Curve (AUPRC), as it is appropriate for imbalanced datasets. About 9\% of pixels in the dataset are labeled as landslides, which makes the baseline AUPRC 0.09. 

A U-Net~\cite{RonnebergerUNetConvolutionalNetworks2015} model with a ResNet-50 encoder (48,982,754 trainable parameters) and varying number of input channels is trained for 100 epochs using an Adam optimizer (weight decay=0.0001) and a ReduceLROnPlateau\footnote{\url{https://pytorch.org/docs/stable/generated/torch.optim.lr_scheduler.ReduceLROnPlateau.html}} learning rate schedule (factor=0.1, patience=10, threshold=0.0001) with an initial learning rate of 0.01.
We use a standard cross-entropy loss to compare the model outputs to the landslide segmentation masks. VV and VH channels are log-transformed and standardized before serving as input. 

We test different combinations of inputs channels (shown in Figure \ref{fig:1}) to identify the most informative features and their combinations for the landslide detection task. 
Along with SAR polarimetry data before and after the event we use as input DEM-derived features (elevation, slope, aspect and curvature). When using multiple timesteps of SAR inputs, we take the pixel-wise mean value.
This means that using a combination of VV/VH polarimetry bands results in 4 input channels (VV before, VH before, VV before, VV after). 
We test how many timesteps/satellite passes before and after the event need to be aggregated to obtain reliable landslide segmentation predictions.

We run experiments with the following inputs:
\begin{itemize}
    \item \textit{SAR (VV)}: Only VV band before and after the earthquake (total 2 input channels).
    \item \textit{SAR (VH)}: Only VH band  before and after the earthquake (total 2 input channels).
    \item \textit{SAR (VV,VH)}: Only SAR bands (VV and VH) are used before and after the earthquake (total 4 input channels).
    \item \textit{SAR+DEM}: SAR bands (VV and VH) before and after the earthquake, plus DEM-derived data (elevation, slope, aspect, curvature) are used (total 8 input channels).
\end{itemize}

For each input configuration, we train using 1, 2, 3 or 4 time steps (average pixel-wise into 1 channel) before and after the event. Notice that regardless the number of time steps we use, the number of SAR channels remains constant as we take the pixel-wise mean value across all time steps.

All models were trained on 1 NVIDIA Tesla V100 GPUs on the Google Cloud Platform. Training one model requires approximately 0.25 GPU hours.

% Each data product is fed into the model as a separate input channel.
% For time-dependent data products, we input the time-step before and the time-step after the event as separate input channels.
% When using multiple timesteps, we average the channels in the timestep dimension before and after the event.

% We run experiments for different combinations of polarimetric channels and data derived from a Digitial Elevation Model (DEM), i.e. slope, aspect, and curvature. 

%We further validation and test sets that are taken from geographically distinct areas within the datasets.
%This ensures that there is no leakage of information from the training set into the test or validation sets. 

%Include code, data and instructions needed to reproduce the main experimental results

% \subsubsection{Compute resources}

%Include total amount of compute and the type of resources used (e.g., type of GPUs, internal cluster, or cloud provider)

\section{Results}

\begin{figure}[htbp]
  \centering
  \includegraphics[width=0.55\linewidth]{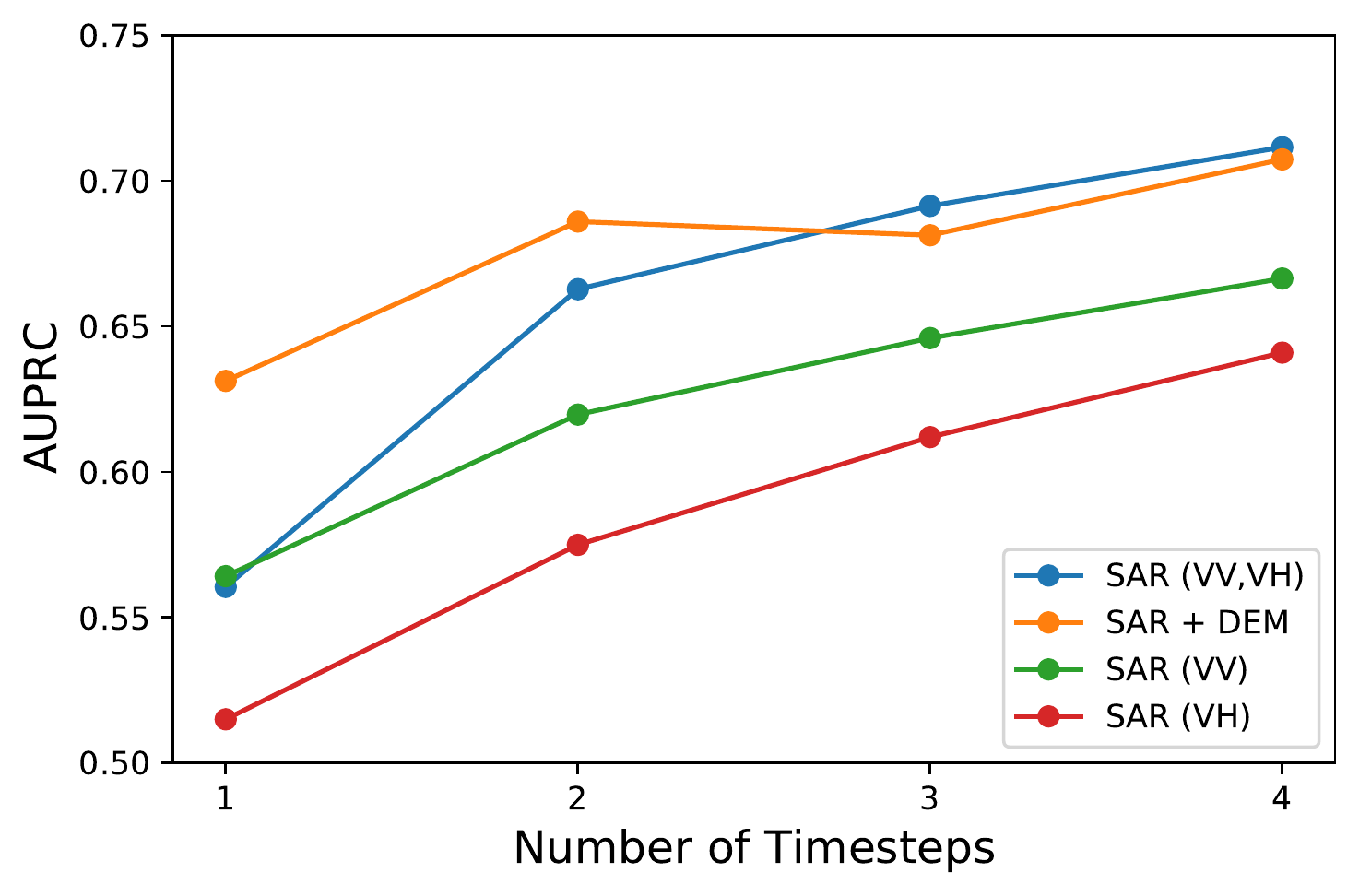}
  \caption{
    Test Area under the Precision-Recall Curve (AUPRC) metrics for different experiments.
    The x-axis shows the number of SAR timesteps used, while the y-axis shows the AUPRC metric.
    The four lines represent experiments using various combinations of Synthetic Aperture Radar (SAR) and Digital Elevation Model (DEM) inputs, from bottom to top (red: VH channel only; green: VV channel only; orange: SAR VV+VH channels + DEM; blue: SAR VV+VH channels with no DEM).
  }
  \label{fig:4_results}
\end{figure}

Results of our input ablation study are shown in Fig~\ref{fig:4_results}.
We find that increasing the number of time steps increases the performance of the model for all input configurations. 
VV is more informative than VH for landslide detection in Hokkaido, and the combination of VV and VH achieves the best results.
Combining the SAR inputs with DEM-derived products, we can achieve better AUPRC when only 1 or 2 satellite overpasses are available. 
This shows that the addition of DEM-derived data can facilitate early detection in crisis-management scenarios. There is no evident gain from adding DEM-derived channels when more than 2 SAR timesteps are available.

It is important to note that a random model achieves a baseline AUPRC of 0.09 and a model trained with just the DEM and DEM-derived features achieves a maximum AUPRC of 0.17.
This demonstrates the value added by SAR data and the complementarity of SAR and DEM input channels.

\begin{figure}[htbp]
  \centering
  \includegraphics[width=0.8\linewidth]{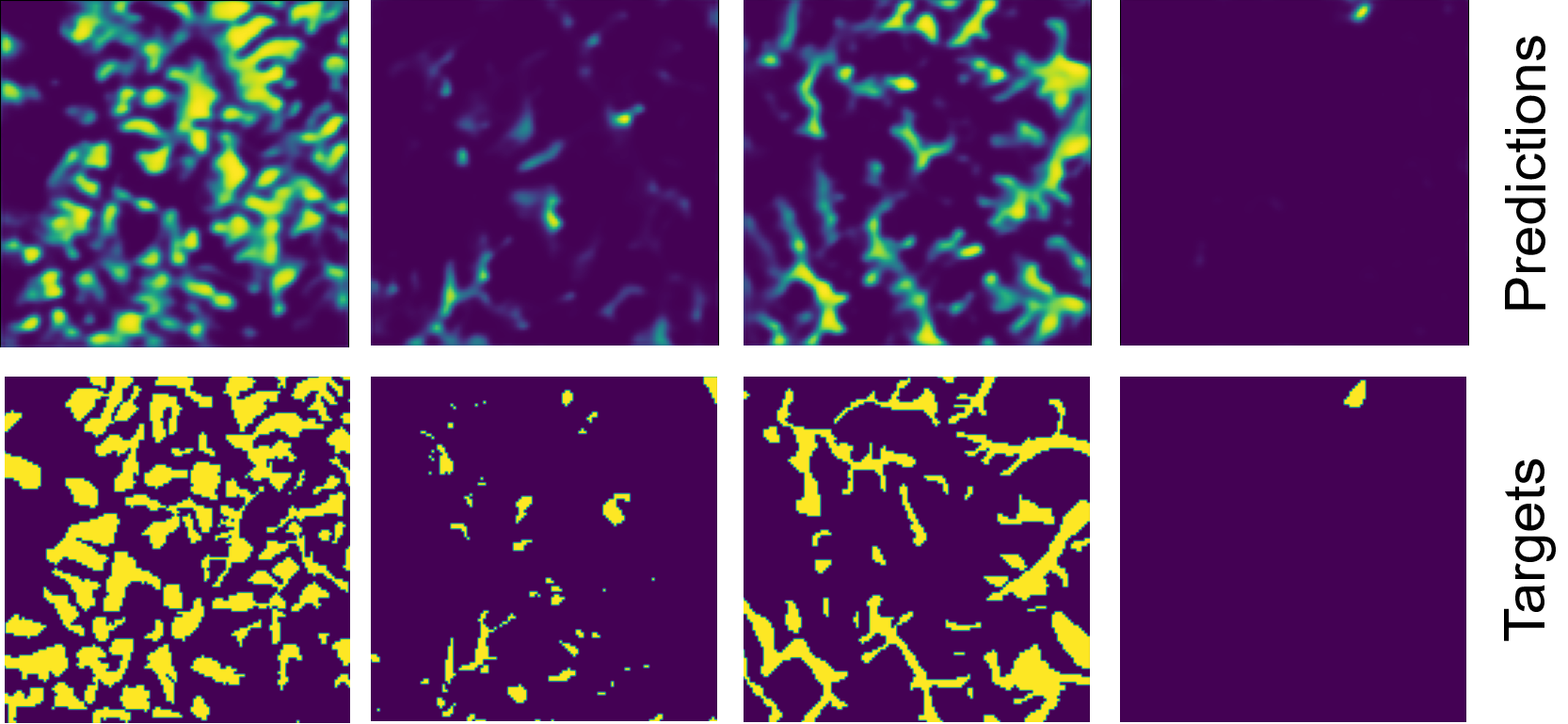}
  \caption{
    Examples of predicted landslide maps (top row) versus groundtruth target maps (bottom row) from the test set.
    Background (no landslide) areas in purple, landslide affected areas in yellow.
  }
  \label{fig:5_results_maps}
\end{figure}

\section{Conclusion}
Rapid and automated disaster assessment is becoming more crucial than ever as climate change increases both the severity and frequency of weather-induced disasters.
Here, we have demonstrated the feasibility of using deep learning to detect landslides from complex Synthetic Aperture Radar data and measure the trade offs of time series data for rapid response scenarios.
Both the scalability of deep learning methods and the fact that SAR measurements are not obstructed by weather or night time conditions have the potential to significantly shorten assessment times.

To further foster research in this direction, and help bridge the barriers for computer vision specialists to work with SAR data, we make public the analysis-ready data cubes for landslide detection \cite{boehm_vanessa_2022_7248056}.

Future work for this use case could include measuring the contribution on SAR interferometry, self-supervised pretraining on a subset of the datacube and understanding how this model could be applied to other regions of the world with different terrain, vegetation coverage and geological characteristics.

% \section{Acknowledgements}
% This work was funded by the FDL/NASA 2022 research sprint. For their invaluable insights we are indebt to Aaron Piña (NASA SMD), Eric Fielding, Erica Podest and Alex Handwerger (NASA JPL), John Stock, Jason Stocker and Corina Cerovski-Darriau (USGS), Forrest Williams and Frank Mayer (Alaska Satellite Facility), Ioannis Papoutsis (National Greek Observatory) and Ronny Haensch (DLR).
\begin{ack}
This work has been enabled by the Frontier Development Lab Program (FDL).
FDL is a collaboration between SETI Institute and Trillium Technologies Inc., in partnership with Department of Energy (DOE) , National Aeronautics and Space Administration (NASA), and U.S. Geological Survey (USGS).
The material is based upon work supported by NASA under award No(s) NNX14AT27A.
Any opinions, findings, and conclusions or recommendations expressed in this material are those of the authors and do not necessarily reflect the views of the National Aeronautics and Space Administration.

We thank Aaron Pina, our contact from NASA who helped shape this research.
He also facilitated communication with experts from NASA and ASF, namely Gerald Bawden (NASA), Eric Fielding (NASA), Alex Handwerger (NASA), Erika Podest (NASA), Franz Meyer (ASF), Forrest Williams (ASF).
We also thank our FDL reviewers Brad Neuberg (Planet), Ronny Hänsch (DLR), Erika Podest (NASA), Ioannis Papoutsis (National Observatory of Athens) who provided great feedback on specificities of ML for SAR.
% Use unnumbered first-level headings for the acknowledgments. All acknowledgments
% More information about this disclosure can be found at: \url{https://neurips.cc/Conferences/2020/PaperInformation/FundingDisclosure}.

% Do {\bf not} include this section in the anonymized submission, only in the final paper. You can use the \texttt{ack} environment provided in the style file to automatically hide this section in the anonymized submission.
\end{ack}

\bibliography{references,ML_references}

\newpage
\appendix

\section{Datacube creation}
\label{app:datacubes}

Earth Observation datacubes are analysis-ready data \cite{DwyerAnalysisReadyData2018,SternPangeoForgeCrowdsourcing2022}, an intermediate product flexible enough to be used by both remote sensing scientists looking at time-series data on a geographic region, and by machine learning experts who can ingest the data as tensors into neural network models.

Multiple streams of the data processing feed into the datacubes.
Given a geographical area of interest (AOI) and a time range, we perform a spatiotemporal query to several SAR satellite data providers that return a list of raw Sentinel-1 granules matching the search criteria.
SAR data is composed of amplitude and phase information and we handle them separately in our data pipeline.

\begin{figure}[htbp]
  \centering
  \includegraphics[width=1.0\linewidth]{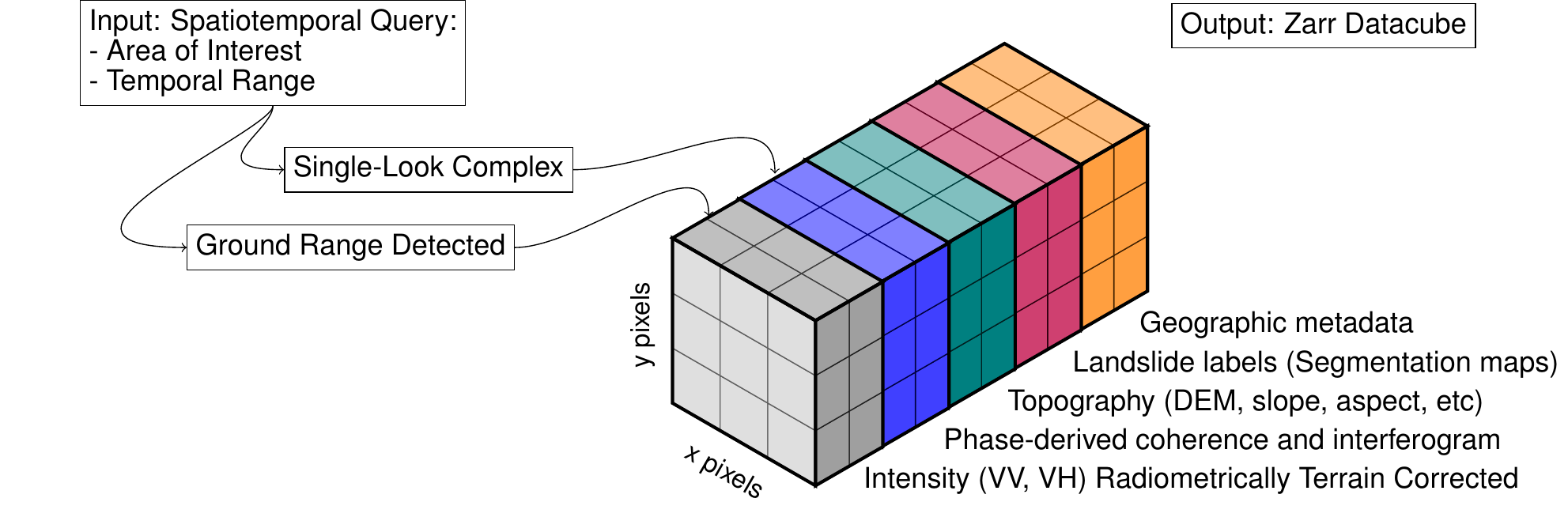}
  \caption{
    Datacube containing Synthetic Aperture Radar (SAR) data and other geographic layers.
  }
  \label{fig:a1}
\end{figure}

For amplitude-based SAR products, we directly obtain Sentinel-1 Radiometrically Terrain Corrected (RTC) data hosted on Microsoft Planetary Computer\footnote{\url{https://planetarycomputer.microsoft.com/dataset/sentinel-1-rtc}} which has been processed from Ground Range Detected (GRD) data.
The RTC data is stacked both in the polarimetric channels and time to produce an xarray dataset \cite{HoyerXarrayNDLabeled2017} with dimensions (time, y, x), and two polarimetric directions, VV and VH.

For the phase-based SAR product, we download the Sentinel-1 Single-Look Complex (SLC) product from the Alaska Satellite Facility (ASF).
SLC data is further processed by using the phase information of two consecutive time steps (timepair, with a start and end date) to create interferogram and coherence maps with interferometry.
SAR interferometry is used to measure the surface deformation with differences in phase of two SAR images over a region of interest and coherence is an indicator for agreement between the two phases.
For these processing step, we used the InSAR Scientific Computing Environment (ISCE, version 2) software package \cite{RosenInSARScientificComputing2012,RosenInSARScientificComputing2018} developed at the NASA Jet Propulsion Laboratory, California Institute of Technology.
The topsApp program of the software is used on the interferometric wide (IW) swath mode SLC products with azimuth looks = 1 and range looks = 5 for processing of the interferograms.
This includes steps such as computing the baseline using orbit files, co-registration with enhanced spectral diversity, calculating interferograms, geocoding.
Note that no phase unwrapping is performed here.
Thus phase-derived products have dimensions (timepair, y, x) and the data variables are `interferogram' and `coherence'.

Besides the Sentinel-1 SAR data, we include two additional geographical datasets in our data pipeline.
The Shuttle Radar Topography Mission (SRTM) DEM \cite{RabusShuttleRadarTopography2003} is downloaded, and processed to derive slope and aspect data layers.
Landslide vector data for the multiple regions is obtained from multiple sources~\cite{SchulzGeographicInformationSystem2021,MasseyVersionLandslideInventory2021,ZhangCharacteristicsLandslidesTriggered2019,UNOSATLandslideAnalysisMount2022}.
The vector files from these sources are rasterized to the same pixel resolution and spatial extent as the other raster SAR data.

Finally, we assemble all the SAR and non-SAR products into a single datacube.
The datacube itself is initialized from the RTC data with dimensions (timestep, y, x) and data variables VV and VH.
Next, the DEM and landslide labels are appended as new data variables along the y and x dimensions.
Then, the phase-derived products are appended, not just along the y and x dimensions, but also along a new ‘timepair’ dimension containing the first and second dates of the Sentinel-1 scenes used to create the interferogram.
All of these appended products are geographically reprojected and aligned using rioxarray \cite{SnowCortevaRioxarray112022} if necessary to match the RTC template.
The resulting xarray Dataset \cite{HoyerXarrayNDLabeled2017} has dimensions (timepair, timestep, y, x) and is stored in a cloud-friendly Zarr format \cite{MilesZarrdevelopersZarrpython2022}. The datasets can be found in Zenodo \cite{boehm_vanessa_2022_7248056}.

\end{document}